\documentclass[useAMS,usenatbib]{mn2e}
\usepackage{graphicx}
\usepackage{wasysym}
\usepackage{lscape}

\title{Qatar-1b: a hot Jupiter orbiting a metal-rich K dwarf star}
\author[K. Alsubai et~al.]
{
K.\,A.\,Alsubai$^{1}$\thanks{E-mail:kalsubai@qf.org.qa},
N.\,R.\,Parley$^{2}$,
D.\,M.\,Bramich$^{3}$,
R.\,G.\,West$^{4}$,
P.\,M.\,Sorensen$^{5}$,
\newauthor
A.\,Collier Cameron$^{3}$,
D.\,W.\,Latham$^{6}$,
K.\,Horne$^{3}$,
D.\,R.\,Anderson$^{7}$,
D.\,J.\,A.\,Brown$^{2}$,
\newauthor
L.\,A.\,Buchhave$^{8}$,
G.\,A.\,Esquerdo$^{6}$,
M.\,E.\,Everett$^{9}$,
G.\,F\H{u}r\'esz$^{6}$,
C.\,Hellier$^{7}$,
\newauthor
G.\,M.\,Miller$^{2}$,
D.\,Pollacco$^{10}$,
S.\,N.\,Quinn$^{6}$,
J.\,C.\,Smith$^{11}$,
R.\,P.\,Stefanik$^{6}$,
\newauthor
and
A.\,Szentgyorgyi$^{6}$.
\\
$^{1}$Qatar Foundation, P.O.BOX 5825, Doha, Qatar.
\\
$^{2}$SUPA, School of Physics and Astronomy, University of St Andrews, 
North Haugh, St Andrews, Fife KY16~9SS, UK.
\\
$^{3}$European Southern Observatory, Karl-Schwarzschild-Stra$\beta$e 2, 
85748 Garching bei M\"{u}nchen, Germany
\\
$^{4}$Department of Physics and Astronomy, University of Leicester, 
LE1~7RH, UK.
\\
$^{5}$Nordic Optical Telescope, Apartado 474, E-38700 Santa Cruz de la Palma,
Santa Cruz de Tenerife, Spain
\\
$^{6}$Harvard-Smithsonian Center for Astrophysics, 60 Garden Street, 
Cambridge, MA 02138, USA.
\\
$^{7}$Astrophysics Group, Keele University, Staffordshire, ST5 5BG, 
UK.
\\
$^{8}$Niels Bohr Institute, Copenhagen University, DK-2100 Copenhagen, 
Denmark.
\\
$^{9}$Planetary Science Institute, 1700 East Fort Lowell Road, Suite 
106, Tucson, AZ 85719, USA.
\\
$^{10}$Astrophysics Research Centre, School of Mathematics \&\ Physics,  
Queens University, University Road, Belfast, BT7~1NN, UK.
\\
$^{11}$Hidden Loft Observatory, Tucson, AZ 85755, USA.
\\
}

\begin{document}

\date{Accepted 0000 Dec 00. 
Received 0000 Dec 00; in original form 0000 October 00}

\pagerange{\pageref{firstpage}--\pageref{lastpage}} \pubyear{2010}

\maketitle

\label{firstpage}

 \begin{abstract}
 We report the discovery and initial characterisation of Qatar-1b, a hot 
Jupiter orbiting a metal-rich K dwarf star, the first planet discovered 
by the Alsubai Project exoplanet transit survey. We describe the 
strategy used to select candidate transiting planets from 
photometry generated by the Alsubai Project instrument. We examine
the rate of astrophysical and other false positives found during the 
spectroscopic reconnaissance of the initial batch of candidates. A simultaneous
fit to the follow-up radial velocities and photometry of Qatar-1b yield 
a planetary mass of $1.09 \pm 0.08 M_{\rm J}$ and a radius of $1.16\pm 0.05 R_{\rm J}$.
The orbital period and separation are 1.420033 days and 0.0234 AU for an orbit
assumed to be circular.
The stellar density, effective temperature and rotation rate indicate an age 
greater than 4 Gyr for the system.

 \end{abstract}


\section{Introduction}

Transiting extrasolar planets are important because measurements of the planetary transits as well as the stellar reflex velocity provide both the mass and radius, and hence the density of the planet. In contrast to the relatively tight mass-radius relationship of main-sequence stars, the hot Jupiters found in transit surveys exhibit a wide range of radii at each mass, thus additional parameters affect the radii of close-in gas giants. With over 100 transiting extrasolar planets now securely characterised\footnote{see http://exoplanet.eu} statistics are beginning to support comparative studies to unravel the factors that determine their radii and orbit parameters \citep{2009A&A...501.1161M}. Toward this goal it is important to extend the statistics of hot Jupiters to smaller planets.

The Kepler satellite has delivered over 700 transiting objects, providing good statistics on the orbital periods and radii of transiting bodies \citep{2010arXiv1006.2799B}. At typical distances 300-1000~pc, follow-up spectroscopic studies for high-precision radial velocities to measure the stellar wobble and hence the masses of the smaller transiting bodies presents a considerable challenge for present-generation instruments. The wide-angle ground-based exoplanet transit surveys, in contrast to Kepler, survey brighter stars typically within 100~pc. For these relatively nearby stars, follow-up observations are considerably easier.

The Alsubai Project\footnote{http://www.alsubaiproject.org/default.aspx} \citep{alsubai2011project} has initiated a wide-field transit search programme deploying initially a 5-camera CCD imaging system designed to go deeper than most current wide-angle survey systems such as SuperWASP\footnote{see http://www.superwasp.org/}, HATNet\footnote{http://www.hatnet.hu/}, TrES\footnote{http://dl.dropbox.com/u/502281/Sites/solas/tres/tres.html} and XO\footnote{http://www-int.stsci.edu/$\sim$pmcc/xo/}. The Alsubai Project's first site in New Mexico was chosen to complement the SuperWASP sites in the Canary Islands and South Africa, where suites of eight 200mm f/2.0 Canon lenses image $16^\circ\times32^\circ$ degree fields at 15~arcsec~pix$^{-1}$. Combining Alsubai and SuperWASP data should increase the exoplanet discovery rate by enabling multiple transits of candidate systems to be obtained more quickly. The Alsubai camera's higher angular resolution and larger aperture should also extend the search for transiting planets to fainter stars with smaller radii, and hence to the discovery of smaller transiting planets.

This paper reports the discovery and initial characterisation of the first confirmed transiting exoplanet to emerge from the Alsubai Project exoplanet transit survey, orbiting the star 3UC311-087990 (Qatar-1, $\alpha_{2000}=20^h13^m31^s.61$, $\delta_{2000}=+65^\circ09'43''.4$). The survey observations and candidate selection procedures that led to the discovery of Qatar-1b are presented in Section~\ref{sec:obs}, together with our analysis of the stellar spectrum and follow-up photometry. We discuss in Section~\ref{sec:star} our determination of the stellar and planetary system parameters, leading to the summary and conclusions in Section~\ref{sec:conclude}.

\section{Observations}
\label{sec:obs}

The Alsubai camera system images an $11^\circ\times11^\circ$ field of view simultaneously at two pixel scales. A single 200mm f/2.0 Canon lens covers the full $11^\circ\times11^\circ$ degree field at 9.26~arcsec~pix$^{-1}$. Four 400mm f/2.8 Canon lenses each cover $5.5^\circ\times5.5^\circ$ fields arranged in a $2\times2$ mosaic to cover the same $11^\circ\times11^\circ$ field. All cameras use FL1 ProLine PL6801 KAF-1680E 4K$\times$4K CCD detectors. The robotic mount cycles through four pointings, taking 100s exposures with the 400mm cameras and 60s exposures with the 200mm camera, thereby covering a $\sim400$~square degree field with a cadence of 8~minutes. The 400mm lenses target stars between $V=11$ and 15  mag with 100-s exposures, while the 200mm lens covers the magnitude range  from $V=8$ to 12 mag with 60-s exposures. We thus obtain photometry of all stars in the field in the range from $V=8$ to 15 mag.

The data are reduced at the University of St Andrews using pipeline software based on the image-subtraction algorithm of \citet{2008MNRAS.386L..77B}. A detailed description of the pipeline is given by \citet{alsubai2011project}. The pipeline data products are ingested into a data archive at the University of Leicester, which uses the same architecture as the WASP archive \citep{2006PASP..118.1407P}.

\subsection{Discovery photometry}

An automated transit search was conducted on the archive data using the box 
least-squares (BLS)  algorithm of \citet{2002A&A...391..369K} as modified for the 
SuperWASP project by \citet{2006MNRAS.373..799C}. Systematic patterns of 
correlated noise were modelled and removed from the archive light curves using 
a combination of the {\sc SysRem} algorithm of \citet{2005MNRAS.356.1466T} 
and the Trend Filtering Algorithm (TFA) of \citet{2005MNRAS.356..557K}.
The light curves of all candidates were subjected to the candidate screening tests 
described by \citet{2007MNRAS.380.1230C} to ensure that the depths and durations
of the transits were consistent with expectation for objects of planetary 
dimensions transiting main-sequence stars. In cases where the same star had been 
observed by the SuperWASP survey, we ran periodogram tests to seek evidence of 
a transit signal in the SuperWASP data at the same period. 

The star 3UC311-087990 was found to exhibit transit-like events at 
1.42-day intervals in three fields of the Alsubai Project instrument. 
Field 073723+532023 was observed 2239 times from 2010 Jun 16 to Sep 7, 
while fields 074307+520936 and 074440+552202 were both observed on 2916 
occasions between 2010 Jun 16 and Sep 24. These fields are to the north 
of the declination limit of the SuperWASP survey, so no SuperWASP light 
curve was available. In all three fields, 3UC311-087990 exhibited a clear 
transit signal with signal detection efficiencies (as defined by \citealt{2002A&A...391..369K})
$SDE=11$, 22 and 16 respectively for the three fields. The corresponding 
signal-to-red noise ratios were $S_{\rm red}=11$, 12 and 13 using the 
definition of \citet{2006MNRAS.373..799C}.

The transit duration and $J-H$ colour of 3UC311-087990 were found to be 
consistent with the radius and mass of a 
main-sequence K dwarf host. For such a star, the 0.02-mag transit depth 
suggests a companion radius close to that of Jupiter.

\begin{figure}
 \includegraphics[width=0.4\textwidth]{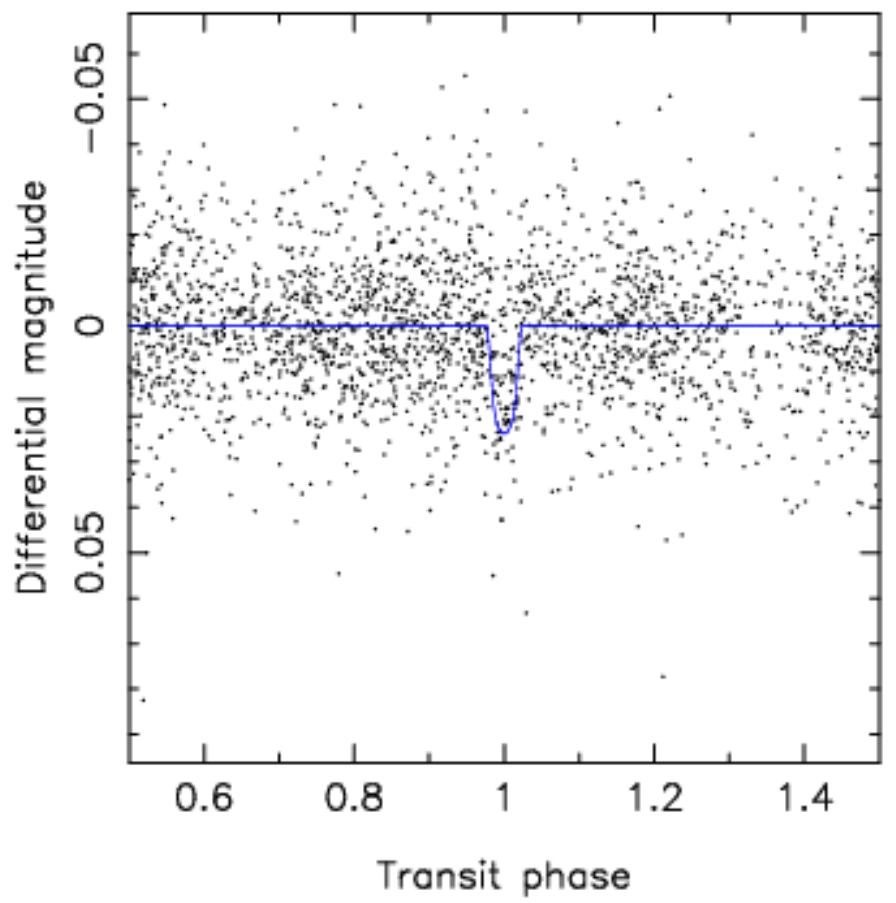}
 \caption{Alsubai discovery light curve of Qatar-1b. The data are 
phase folded using the ephemeris ${\rm HJD}=2455518.4102+1.420033E$.
The solid line represents the best fitting model transit light curve derived from these observations 
and the follow-up radial-velocities and photometry.}
 \label{fig:disclc}
 \end{figure}

\clearpage
\begin{landscape}
\begin{table}
\begin{center}{
  \caption[Alsubai transiting planet candidates.]{Alsubai transiting planet candidates. The columns are labelled as follows:\\
column 1: UCAC3 identification for the target, \\
columns 2 and 3: J2000 Right Ascension and Declination measured by this project, \\
column 4: V magnitude from Alsubai pipeline calibration against UCAC3 magnitude, \\
columns 5 and 6: Photometric period and epoch, \\
columns 7 to 9: Effective temperature, log surface gravity,
and rotational line broadening of the synthetic template spectrum that
gave the best match to the observed spectra, assuming solar metallicity, \\
columns 10 and 11: Absolute radial velocity on the IAU system and rms
variation when two or more measurements were obtained, \\
columns 12 and 13: Number of TRES observations and total exposure time, \\
column 14: Detection summary: A$n$ means the transits were detected by
$n$ of the Alsubai cameras, W1 means the transits were also detected
with the same period by SuperWASP, W0 means that SuperWASP did not
confirm the detection, and no W means the target field was not
observed by SuperWASP, \\
column 15: Disposition of the candidate: \\
D = the spectrum shows two sets of lines;
FA? = a likely photometric false alarm;
FR = a rapidly rotating star for which precise radial velocities are not feasible;
G = a giant, presumably in a blended system with an eclipsing binary;
G? = a likely giant;
H = a hot star;
P? = the velocity variations are small and not inconsistent with a planetary companion;
P! = a confirmed planetary companion;
S = the spectrum is single-lined and shows a large velocity variation due to a stellar companion;
T? = a likely triple system.
  }
  \label{tab:candidates}
  \begin{tabular}{lccccccccrrcrrr}
   \hline
   Candidate & $\alpha_{2000}$ & $\delta_{2000}$ & $V$ & Period & Epoch & $T_{\rm eff}$ & $\log(g)$ & $V_{\rm rot}$ & $V_{\rm rad}$ & $\sigma_{V}$& $N_{\rm obs}$ & ExpT & Field & Type  \\
 \hline
    &  &  & mag & days & HJD & K &  & $\rm km\ s^{-1}$ & $\rm km\ s^{-1}$ & $\rm km\ s^{-1}$ &  & min &  &   \\
   \hline
215-018836 & 04:20:29.30 & 17:03:29.0 & 11.7 & 2.6539 & 2455182.03187 & 5000 & 3.00 &   6 & $  32.050 $ & \ldots & 1 &  30 & A1   & G \\  
210-017759 & 04:23:18.22 & 14:58:19.2 & 14.2 & 1.6343 & 2455170.02498 & 7750 & 5.00 &   6 & $	0.737 $ &  0.860 & 2 & 105 & A1W0 & H \\  
214-019086 & 04:25:10.26 & 16:49:59.9 & 14.8 & 2.1799 & 2455186.89730 & 6250 & 4.50 &  60 & $  -5.823 $ & \ldots & 1 &  80 & A1W1 & FR \\ 
214-019418 & 04:27:58.86 & 16:55:21.8 & 13.3 & 1.2068 & 2455510.83400 & 6125 & 3.75 &   7 & $ -41.841 $ &  0.075 & 2 &  70 & A1W0 & FA? \\
214-019710 & 04:30:22.44 & 16:54:27.1 & 14.0 & 2.3042 & 2455180.03781 & 6250 & 5.00 &  25 & $  -5.879 $ & \ldots & 1 &  45 & A1W0 & \ldots \\
216-021225 & 04:35:04.30 & 17:35:27.9 & 13.3 & 1.2131 & 2455154.58890 & 5875 & 3.75 &   7 & $  13.823 $ &  0.078 & 2 &  54 & A1W0 & FA? \\
212-019540 & 04:35:15.20 & 15:30:57.9 & 12.6 & 5.2482 & 2455155.24280 & 6125 & 3.50 &  45 & $  -1.858 $ &  1.281 & 2 &  80 & A1   & FR \\ 
197-017651 & 04:38:58.36 & 08:13:15.0 & 13.3 & 1.3681 & 2455511.60200 & 6750 & 4.00 &   9 & $  86.048 $ &  0.061 & 3 &  76 & A1W0 & FA? \\
183-015913 & 04:41:57.49 & 01:24:48.7 & 12.5 & 1.5418 & 2455178.86996 & 6250 & 3.50 &  20 & $  64.420 $ & \ldots & 1 &  24 & A1   & D \\  
210-019969 & 04:45:41.07 & 14:43:21.6 & 13.1 & 1.9023 & 2455154.86690 & 5250 & 4.50 &  25 & $  26.303 $ & \ldots & 1 &  24 & A1W0 & D \\  
217-022884 & 04:47:31.64 & 18:29:08.8 & 13.1 & 2.1060 & 2455155.62150 & 6875 & 4.00 &  28 & $  -6.508 $ &  0.434 & 2 &  47 & A1W0 & H \\  
184-018774 & 05:03:22.93 & 01:45:07.2 & 13.4 & 1.8672 & 2455154.95550 & 6750 & 3.50 &  20 & $ -22.110 $ & \ldots & 1 &  30 & A2W1 & D \\  
194-020871 & 05:03:59.84 & 06:48:36.8 & 13.2 & 2.3520 & 2455184.59328 & 6000 & 5.00 &  16 & $ -19.136 $ & \ldots & 1 &  30 & A1W1 & D \\  
197-023150 & 05:09:14.35 & 08:16:52.8 & 12.2 & 4.2357 & 2455156.63930 & 6000 & 4.00 &   8 & $  -8.243 $ &  0.169 & 2 &  60 & A1W0 & FA? \\
199-022618 & 05:09:35.29 & 09:17:43.2 & 14.1 & 1.7068 & 2455182.41670 & 6500 & 4.00 &  30 & $  40.073 $ & 54.218 & 2 &  84 & A1W1 & S \\  
196-023727 & 05:16:02.99 & 07:37:03.4 & 13.7 & 1.8731 & 2455154.93060 & 4750 & 4.50 &   1 & $ -11.494 $ &  0.118 & 3 & 150 & A1W0 & FA? \\
205-023986 & 05:25:06.47 & 12:04:33.7 & 11.3 & 5.2372 & 2455175.61072 & 5000 & 3.00 &   4 & $  27.900 $ & \ldots & 1 &  15 & A1   & G \\  
190-024393 & 05:28:13.25 & 04:42:24.4 & 13.3 & 1.3150 & 2455180.88287 & 5417 & 4.33 &   3 & $  29.785 $ &  0.267 & 6 & 289 & A1W0 & FA? \\
210-028815 & 05:30:41.80 & 14:54:49.0 & 12.9 & 2.8459 & 2455155.23150 & 6000 & 3.75 &   9 & $  28.885 $ &  0.121 & 2 &  87 & A1   & P? \\ 
308-105347 & 18:51:21.19 & 63:59:02.5 & 12.8 & 1.5758 & 2455365.00960 & 6500 & 3.50 &  45 & $-113.203 $ & \ldots & 1 &  30 & A2W1 & D \\  
284-137549 & 18:59:30.39 & 51:34:50.0 & 12.4 & 2.4046 & 2455365.53430 & 6000 & 3.00 &  45 & $  19.963 $ & \ldots & 1 &  24 & A1W1 & T? \\ 
304-111233 & 19:22:52.65 & 61:53:48.6 & 13.3 & 2.0262 & 2455510.61000 & 5250 & 4.50 & 110 & $ -14.707 $ & \ldots & 1 &  30 & A2W0 & FR \\ 
309-100007 & 19:25:15.34 & 64:00:17.9 & 13.2 & 2.1804 & 2455407.44653 & 5625 & 4.38 &   3 & $ -36.859 $ &  0.161 & 4 & 175 & A1W0 & FA? \\
301-128216 & 19:32:07.33 & 60:27:07.0 & 13.7 & 1.8227 & 2455364.68300 & 5250 & 4.00 &   2 & $	1.529 $ &  0.104 & 2 &  74 & A2W0 & P? \\ 
301-128743 & 19:35:42.56 & 60:27:05.4 & 12.5 & 1.0851 & 2455413.40090 & 6083 & 4.00 &   6 & $ -28.514 $ & 13.803 & 3 &  72 & A2W1 & T? \\ 
294-143877 & 19:45:11.25 & 56:35:16.3 & 12.2 & 1.3284 & 2455424.53773 & 6250 & 4.50 &   4 & $ -95.671 $ & \ldots & 1 &  30 & A2W1 & D \\  
311-087990 & 20:13:31.61 & 65:09:43.4 & 12.6 & 1.4201 & 2455407.64900 & 4861 & 4.44 &   2 & $ -37.900 $ &  0.189 & 9 & 472 & A3 & P! \\ 
294-152358 & 20:20:22.74 & 56:32:25.0 & 13.0 & 2.3267 & 2455365.32740 & 5000 & 3.50 &   3 & $ -72.159 $ &  0.126 & 2 &  60 & A1W0 & G? \\ 
   \hline
  \end{tabular}
 } \end{center}
\end{table}
\end{landscape}
\clearpage

\subsection{Spectroscopic reconnaissance}

A list of 28 candidates was provided by the Alsubai Project to the CfA team
in November 2010.  
The usual first step at the Harvard-Smithsonian Center for Astrophysics 
(CfA) for vetting transiting-planet candidates from
wide-angle ground-based photometric surveys is to obtain reconnaissance
spectra with the Tillinghast Reflector Echelle Spectrograph (TRES) on the
1.5-m Tillinghast Reflector at the Fred L. Whipple Observatory operated by
the Smithsonian Astrophysical Observatory (SAO) on Mount Hopkins in Southern
Arizona.  These spectra are used to look for evidence of stellar systems
that are the source of the transit-like light curves (e.g. see \citealt{2009ApJ...704.1107L})
and also to provide refined stellar parameters for the targets.  
An initial spectroscopic reconnaissance was carried out
for all 28 candidates, based on 60 spectroscopic observations with TRES
over a span of 17 nights, using the medium fiber at a spectral resolution $R=44000$.  
A summary of the candidates and the results of
the reconnaissance are reported in Table \ref{tab:candidates}.

For the initial analysis, the observed spectra are correlated against a grid of
synthetic spectra drawn from a library calculated by John Laird using
Kurucz models \citep{1992IAUS..149..225K} and a line list prepared by Jon Morse.  The synthetic
spectra cover a window of 300\AA\ centered near the gravity-sensitive Mg~b
features.  The template that gives the best match to the observed spectrum
is then used to derive the rotational and radial velocities.  To establish
the zero point of the radial velocities on the IAU system, a small
correction is derived using nightly observations of standard stars.

The initial analysis of the TRES spectra against the library of synthetic
spectra provides useful information about the characteristics of the target
star, such as effective temperature, surface gravity, and rotational and
absolute radial velocity, but it only uses a small fraction of the full  390 to 900 nm
spectral range of TRES.  To look for evidence of
low-amplitude orbital motion, we take advantage of the wide wavelength
coverage by correlating the individual observations of a star against a
template derived from observations of the same star, either a single
observation that has especially strong signal-to-noise, or a master observed template
constructed by shifting and co-adding all the observations of the star.

The radial velocities reported in Table~\ref{tab:candidates} are calibrated
using an absolute velocity zero point based on observations of the IAU
radial-velocity standard star HD 182488.  The Kepler team has agreed to use
this as the standard for the reconnaissance spectroscopy of Kepler Objects
of Interest.  We have adopted the velocity of $-21.508$ km s$^{-1}$ as the value on
the IAU system for HD 182488. During the November run we accumulated 
15 strong observations of HD 182488, giving an observed mean velocity on
the TRES native system of $-20.807\pm 0.057$ km~s$^{-1}$, where the error is the
rms residuals from the mean.  This is an offset of $+0.701$ km~s$^{-1}$ from the
adopted IAU velocity for HD 182488. The radial velocities reported in Table 1
have had this offset applied, to bring them as closely as possible
into line with the radial velocity of HD 182488 on the IAU system.

\begin{table}
\caption[]{Relative radial velocities for  Qatar-1.}
\label{tab:rvobs}
\begin{tabular}{@{}lcccc}
\hline\\
BJD & Phase & Radial  velocity & Bisector  span \\
 (d) & & (km s$^{-1}$ )& (km s$^{-1}$)  \\
 \hline\\
2455518.69628 &   0.2014 &  $-0.1297  \pm  0.0605$   &   $0.0068 \pm 0.0443$\\
2455519.59174 &   0.8320 &   $0.3515  \pm  0.0292$	 &  $-0.0227 \pm 0.0131$\\
2455520.63231 &   0.5648 &  $0.2411  \pm  0.0462$	 &   $0.0164 \pm 0.0296$\\
2455521.58409 &   0.2350 &   $-0.0663 \pm   0.0407$  &   $0.0127 \pm 0.0166$\\
2455523.60908 &   0.6610 &   $0.2611  \pm  0.0261$	 &  $-0.0045 \pm 0.0143$\\
2455525.59298 &   0.0581 &    $ 0.0   \pm 0.0221$	 &  $-0.0065 \pm 0.0088$\\
2455526.58711 &   0.7582 &   $0.3618  \pm  0.0458$	 &   $0.0037 \pm 0.0219$\\
2455527.59121 &   0.4653 &    $0.0854 \pm   0.0221$  &  $-0.0179 \pm 0.0128$\\
2455528.58427 &   0.1646 &   $-0.0403 \pm   0.0305$  &   $0.0120 \pm 0.0117$\\
\hline\\
\end{tabular}
\end{table}

When many candidates are being vetted, it is efficient to
schedule the initial observations near times of quadrature, as predicted
from the photometric ephemerides.  In the case of double-lined binaries
that are undergoing grazing eclipses, this guarantees that the velocities
of the two stars will be near their maximum separation, and thus a single
spectrum will be sufficient for rejecting the target if it shows a composite 
spectrum.  If the first observation reveals the spectrum of
just one star, a second observation near the opposite quadrature is
optimum for disclosing orbital motion due to unseen stellar companions.
Of course, this assumes that the photometric ephemeris has the correct
period and is up to date.

Nine of the candidates showed clear evidence of stellar companions.  In
eight cases the initial reconnaissance observation revealed a composite
spectrum with at least two sets of lines, with evidence that three stars
were likely to be involved for two of the targets.  For the ninth case the
spectrum was single-lined, but with a very large change in velocity between
the two observations.  Three of the candidates showed broad lines
corresponding to equatorial rotational velocities of tens of km s$^{-1}$,
which would render impractical the measurement of very precise radial
velocities, and two of the candidates proved to have temperatures
corresponding to late A stars.  In three cases the classification of the
TRES spectra indicated that the targets are giants, presumably bright
stars diluting the light of eclipsing binaries, either in hierarchical
triple systems or accidental alignments.  Seven of the candidates showed no
significant velocity variations, but careful inspection of the available
light curves supported the interpretation that the transit detections were
false alarms.  Two of the candidates with secure transit detections showed
very small velocity variations that could be consistent with orbital motion
due to planetary companions, and one candidate that was observed only once
showed a spectrum suitable for very precise velocity measurements.  These
three targets deserve additional follow up.  Finally, one of the candidates
(3UC311-087990, hereinafter referred to as Qatar-1)
was confirmed as a system with a transiting planet, as described in the
following sections.

 \begin{figure}
 \includegraphics[width=0.5\textwidth]{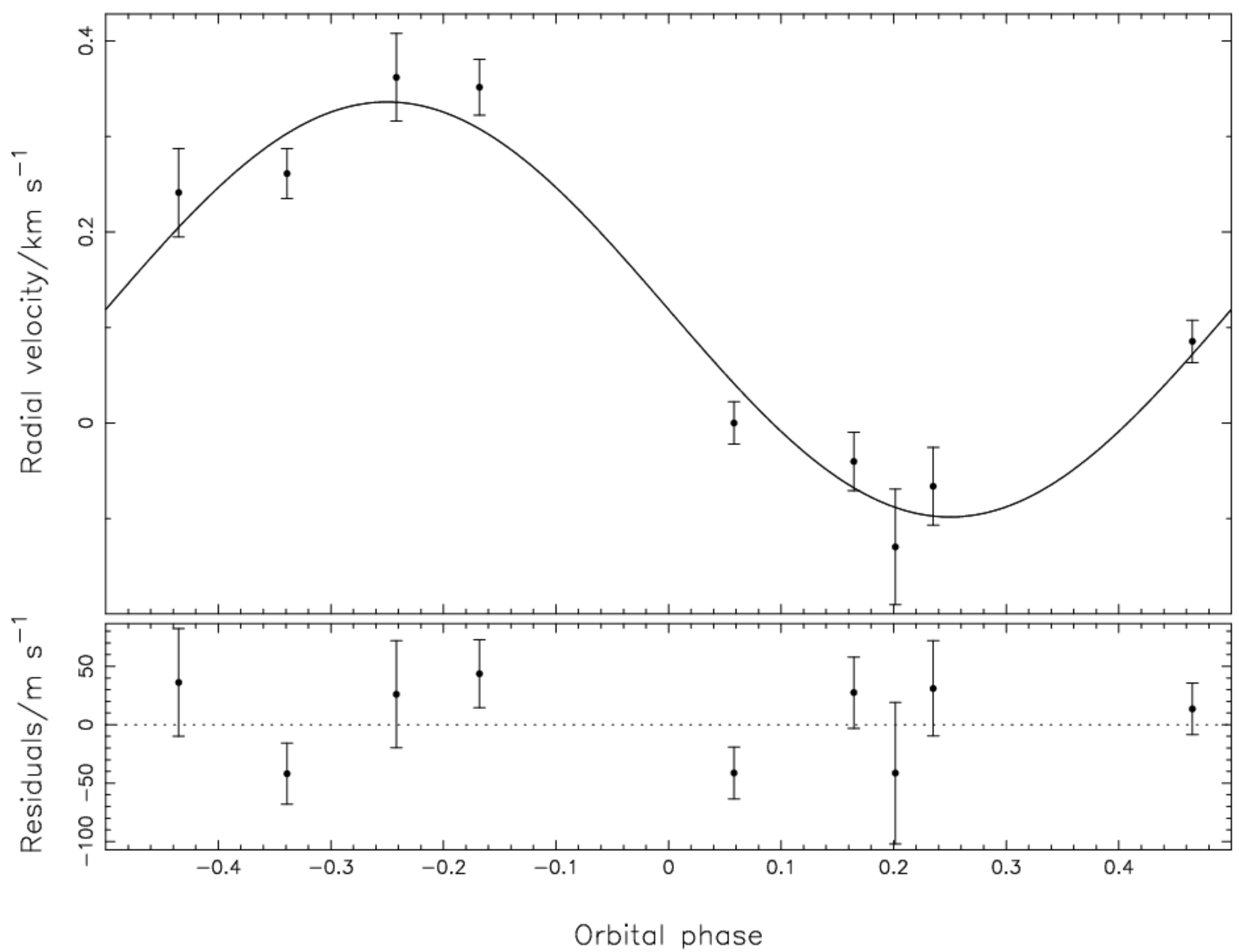}
 \caption{Upper panel: Relative radial-velocity data for Qatar-1, with 
$1~\sigma$ error bars, phase folded on the ephemeris given in 
Table~\ref{tab:params}. The best fitting circular orbit model (solid 
line) is also shown. Lower panel: Residuals of the fit to the best 
fitting circular orbit model. }
 \label{fig:rvorbit}
\end{figure}

\subsection{Radial-velocity  follow-up}

\begin{figure}
\begin{center}{
 \includegraphics[width=0.3\textwidth]{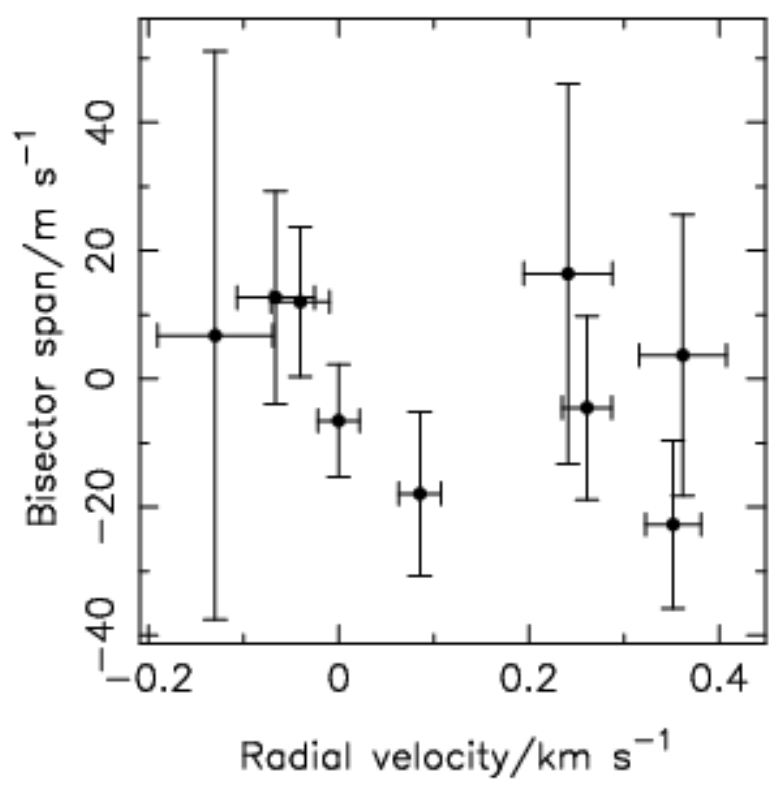}
} \end{center}
 \caption{Line bisector span versus radial velocity for Qatar-1. }
 \label{fig:bsplot}
\end{figure}

The first two radial velocities of Qatar-1,
obtained near opposite quadratures, showed a small but significant
difference consistent with the photometric ephemeris and with the
interpretation of a planetary mass for the companion.  Subsequently this star 
was observed every clear night with a longer exposure time of 54
minutes, with the goal of deriving an orbital solution.  The multi-order
relative radial velocities from all nine observations of Qatar-1 are
reported in Table~\ref{tab:rvobs} and are plotted in Fig.~\ref{fig:rvorbit} together with a circular
orbit, phased to the period and epoch of the photometric ephemeris. No
significant correlation was found between the variation in the line
bisectors and the relative radial velocities, as shown in Fig.~\ref{fig:bsplot}.  Thus
there is no evidence in the TRES spectra that the radial-velocity
variations are due to phenomena other than orbital motion, such as an
unresolved blend with a faint eclipsing binary or spots coupled with
stellar rotation \citep{2001A&A...379..279Q}.

\subsection{Spectroscopic parameters of Qatar-1}

As mentioned in the Introduction, a transiting planet allows us to
determine both the radius and the mass of the planet, if a
spectroscopic orbit for the host star is available to complement the
transit light curve.  This in turn provides key information about the
bulk properties of the planet, such as density.  However, the
planetary mass and radius values are relative to the mass and radius
of the host star, and the accuracy with which the planetary properties
can be determined is often limited by the uncertainties in the
characteristics of the host star.  Nearby stars with accurate
parallaxes have the advantage that the observed luminosity of the star
helps pin down key stellar parameters such as radius and effective
temperature. For more distant stars, such as Qatar-1, the
alternative is to use stellar models together with values for the
effective temperature and metallicity derived from the spectra.

We have used our library of synthetic spectra and a correlation
analysis of our TRES spectra similar to that described by 
\citet{2002AJ....123.1701T}, to derive the following results for Qatar-1:
effective temperature $T_{\rm eff\star} = 4861\pm125\ K$, surface
gravity $\log{g_{\star}} = 4.40\pm0.1$ (log cgs), projected rotational
velocity\footnote{The symbol $I$
represents the inclination of the stellar rotation axis to the line of sight, 
whereas $i$ is used elsewhere in this paper to denote the orbital inclination.} 
$v \sin{I} = 2.1\pm0.8$ km s$^{-1}$ and metallicity $\rm
[Fe/H]=+0.20\pm0.1$\ dex.  Spectroscopic determinations of stellar
surface gravity are notoriously difficult, so it is fortunate that
$\log{g_{\star}}$ can be determined independently from a joint
analysis of the transit light curve and spectroscopic orbit.  For
Qatar-1 that analysis yielded $\log{g_{\star}} = 4.53\pm0.02$,
as described in the next section.

The stellar parameters adopted for the host star are listed in Table~\ref{tab:params}
together with catalogue magnitudes and additional stellar dimensions obtained by fitting
a model to the transit profiles and spectroscopic orbit.

\begin{table}
\caption[]{Stellar parameters for Qatar-1 derived from spectroscopic
reconnaissance, photometric catalogues and model fitting. }
\label{tab:params}
\begin{tabular}{@{}lcc}
Spectroscopic parameter & Value & Source \\
$T_{\rm eff\star}$ (K) & $4861\pm 125$ K & TRES\\
\noindent[Fe/H] & $0.20\pm 0.10$ & TRES\\
$v\sin I$ (km s$^{-1}$) & $2.1\pm 0.8$ & TRES\\
$\gamma_{\rm RV}$ (km s$^{-1}$) & $-37.835\pm 0.063$ &  TRES \\ 

 & & \\
\hline\\
Photometric parameter & Value & Source \\
$V$ (mag) & $12.843\pm 0.137$& TASS4\\
$J$ (mag) & $10.999\pm 0.021$& 2MASS\\
$H$ (mag) & $10.527\pm 0.019$ & 2MASS\\
$K_s$ (mag) & $10.409\pm 0.017$ & 2MASS\\
& & \\
\hline\\
Model parameter & Value & Source \\
$M_\star$ ($M_\odot$) & $  0.85  \pm  0.03 $ & MCMC \\
$R_\star$ ($R_\odot$) & $  0.823  \pm  0.025 $ & MCMC \\
$\rho_\star$ ($\rho_\odot$) & $1.52\pm 0.12$ & MCMC \\
$\log g$ (cgs) & $4.536\pm 0.024$ & MCMC \\
Age (Gyr) & $>6$& MCMC+YY\\
\hline\\
\end{tabular}
\end{table}

\subsection{Follow-up photometry}

A full transit of the planet Qatar-1b was observable from high northern 
latitudes on the night of 2010 Nov 27. $R$-band photometry was 
obtained of the ingress with the CCD camera on the 0.95m James Gregory 
Telescope (JGT) located at St Andrews, Scotland, during the transit of 
2010 Nov 27. A total of 25 180-s exposures was obtained in 
clear conditions, but the sequence was terminated early by snow clouds. 
The egress of the same Nov 27 transit was observed in clear 
conditions with the 60-cm telescope and CCD camera of the University of 
Keele. A sequence of 535 20-s $R$-band 
measurements was obtained.

Using the refined ephemeris from these 
observations we identified an opportunity to observe a complete transit 
using the KeplerCam CCD on the FLWO 1.2m telescope on the evening of 2010 
Dec 2. The transit was observed using 90-s exposures in the Sloan $i$ filter; the 
resulting high-quality photometry was decorrelated against external 
parameters as described by \citet{2010ApJ...710.1724B}. 
A third transit was observed in its entirety on 2010 Dec 7, again using the
JGT with an $R$-band filter.

\begin{figure}
 \includegraphics[width=0.5\textwidth]{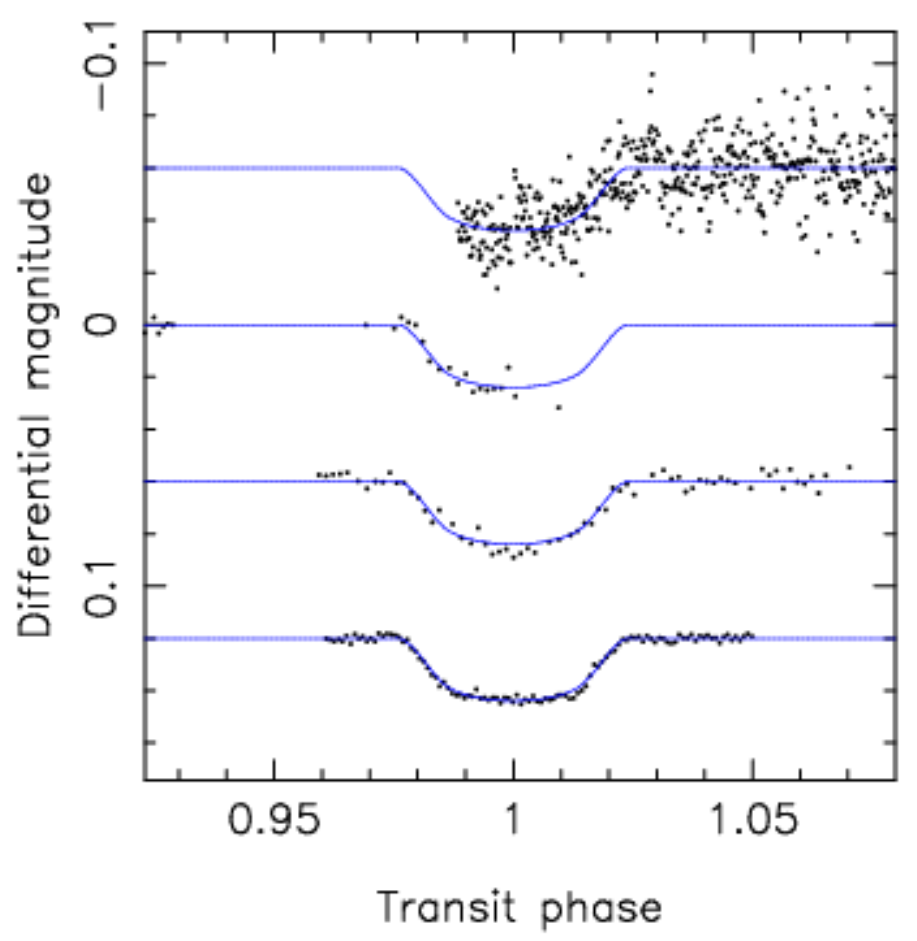}
 \caption{Photometric follow up light curves, offset from each other by 
an arbitrary amount for clarity. From top to bottom:
$R$-band photometry of the planetary egress obtained using the 
Keele 60-cm telescope, 2010 Nov 27;
$R$-band photometry of the Nov 27 planetary ingress obtained using the 
0.95m JGT;
$R$-band JGT photometry of the Dec 7 transit;
Sloan $i$-band photometry of the Dec 2 planetary transit obtained 
using KeplerCam on the 1.2-m telescope at FLWO. All data have been phase folded on the 
ephemeris given in Table~\ref{tab:params}. The best fitting model 
transit light curve is overplotted in all three cases.}
 \label{fig:followphot}
\end{figure}

The four follow-up light curves are shown, together with the best-fitting 
model, in Fig.~\ref{fig:followphot}. The transit is seen to be slightly more than 0.02
magnitudes deep in both wavelength bands. The $i$-band light curve in particular shows 
four well-defined contacts but rather lengthy ingress and egress phases, 
suggesting a moderately high impact parameter.

\section{Stellar and planetary dimensions}
\label{sec:star}

The dimensions of the planet and its host star were determined from a 
simultaneous model fit to the radial velocities and the combined 
photometry from the Alsubai Project cameras and follow-up transit 
observations. The transit light curve was modelled using the formulation 
of \citet{2002ApJ...580L.171M} in the small-planet approximation. A 
4-coefficient nonlinear limb darkening model was used, employing fixed 
coefficients appropriate to the $R$ band for the Alsubai, JGT and Keele 
photometry, and to the Sloan $i$ band for the KeplerCam photometry.  
These were interpolated to the appropriate effective temperature and 
metallicity from the tabulation of \citet{2004A&A...428.1001C}.

The parameter optimization was performed using the current version of 
the Markov-Chain Monte Carlo (MCMC) code described by \citet{2007MNRAS.380.1230C}
and \citet{2008MNRAS.385.1576P}. The transit light curve is modelled in terms of 
the epoch $T_0$ of mid-transit, the orbital period $P$, the ratio of radii 
$d=(R_p/R_\star)^2$, the approximate duration $t_T$ of the transit from
initial to final contact and the impact parameter $b=a\cos i/R_\star$. The radial-velocity
orbit is defined by the stellar orbital velocity semi-amplitude $K_\star$ and the  offset
$\Delta\gamma$ of the centre-of-mass velocity from the zero-point of the relative 
velocities listed in Table~\ref{tab:rvobs}. Where the eccentricity is allowed to 
float, the two additional fitting parameters $e\cos\omega$ and $e\sin\omega$
are introduced, as recommended by \citet{2006ApJ...642..505F}.

 \begin{table*}
 \caption{System parameters and $1\sigma$ error limits derived from the MCMC analysis. Although for reasons given in the text we adopt the circular orbit solution, we include the eccentric solution here to show its influence on other system parameters.}
 \label{tab:mcmc}
 \begin{tabular}{lcccc}
 \hline \\
 Parameter & Symbol & Circular & Eccentric & Units \\
 \hline \\
 Transit epoch & $T_0$ & $  5518.4102   \pm  0.0002$ & $5518.4103 \pm  0.0003$ & days \\ 
 Orbital period & $P$ & $  1.420033   \pm  0.000016$ & $  1.420033   \pm  0.000015$ & days \\ 
 Planet/star area ratio & $(R_p/R_*)^2$ & $  0.02117 \pm  0.00045$ & $  0.02114 \pm  0.00045$ & \\ 
 Transit duration & $t_T$ & $  0.06716 \pm  0.00077$ & $  0.06632 \pm  0.00095$ & days \\ 
 Impact parameter & $b$ & $  0.696^{+  0.021 }_{-  0.024 }$ & $  0.695^{+  0.020 }_{-  0.024 }$ & $R_*$ \\ 
 Stellar reflex velocity & $K_1$ & $  0.218^{+  0.015 }_{-  0.016 }$ & $  0.220^{+  0.015 }_{-  0.016 }$ & km s$^{-1}$ \\ 
 Centre-of-mass velocity offset & $\Delta\gamma$ & $  0.118794^{+  0.000052}_{-  0.000053}$ & $  0.1336^{+  0.0081}_{-  0.0080}$ &  km s$^{-1}$ \\ 
 Orbital eccentricity & $e$ &   0.0 (fixed)  & $0.24^{ +  0.10}_{ -  0.12}$  & \\ 
 Longitude of periastron & $\omega$ & \ldots & $84.6 ^{+  11.9 }_{-10.1}$& degrees \\ 
 Orbital inclination & $i$ & $  83.47 ^{+  0.40  }_{-  0.36  }$ & $  79.4 ^{+ 2.3  }_{-  2.9  }$ & degrees \\ 
 Orbital semi-major axis & $a$ & $  0.02343 ^{+  0.00026}_{-  0.00025}$ & $  0.02363 ^{+  0.00030}_{-  0.00029}$ & AU \\ 
 Planet radius & $R_p$ & $  1.164   \pm  0.045 $ & $  1.47   \pm  0.16 $ & $R_{\rm J}$ \\ 
 Planet mass & $M_p$ & $  1.090   ^{+  0.084 }_{-  0.081 }$ & $  1.132   ^{+  0.096 }_{-  0.087 }$ & $M_{\rm J}$ \\ 
 Planet surface gravity & $\log g_p$ & $  3.265   ^{+  0.044 }_{-  0.045 }$ & $  3.078   ^{+  0.091 }_{-  0.076 }$ & [cgs] \\ 
 Planet density & $\rho_p$ & $  0.690  ^{+  0.098  }_{-  0.084 }$ & $  0.355  ^{+  0.139  }_{-  0.086 }$ & $\rho_{\rm J}$ \\ 
 Planet temperature & $T_{eq}$ & $  1399 \pm  42$ & $  1564 \pm  94$ & K \\ 
 \hline\\
\end{tabular}
\end{table*}

The linear scale of the system depends on the orbital separation $a$, which 
through Kepler's third law depends on the stellar mass $M_\star$. The stellar
mass was estimated at each step in the Markov chain as a function of  the effective
temperature, metallicity and density of the star \citep{2010A&A...516A..33E}. The effective 
temperature and metallicity were treated  as additional MCMC model parameters, 
constrained by Gaussian priors with mean values and variances derived directly
from the stellar spectra, as listed in Table~\ref{tab:params}.

A model fit for an eccentric orbit yields an orbital eccentricity $e=0.23\pm 0.11$. 
The uncertainty in the eccentricity and the orientation of the orbit yields a more
highly inflated and uncertain value for the stellar radius, $R_\star = 1.04\pm 0.11 R_\odot$.
The planet's radius and density are similarly affected.
The stellar mass increases to $0.87\pm 0.03 M_\odot$. A star of this mass would have to be
among the oldest in the galactic disk population to have evolved to such a large radius.
A more likely explanation is that the best-fit value of the eccentricity is spurious, and that the
orbit is close to circular. The improvement in the fit resulting from the addition 
of $e\cos\omega$ and $e\sin\omega$ as fitting 
parameters is insufficient to justify adoption of anything 
other than a circular orbit. The $F$-test approach of \citet{1971AJ.....76..544L} 
indicates that there is a 16.8\% probability that the improvement 
in the fit could have arisen by chance if the underlying orbit were circular. 
In the absence of conclusive evidence to the contrary, we adopted the circular 
orbit model.

The orbital and planetary parameters derived from the MCMC model fit
are summarized in Table~\ref{tab:mcmc}.

\section{Discussion and Conclusions}
\label{sec:conclude}

\begin{figure}
 \includegraphics[width=0.4\textwidth]{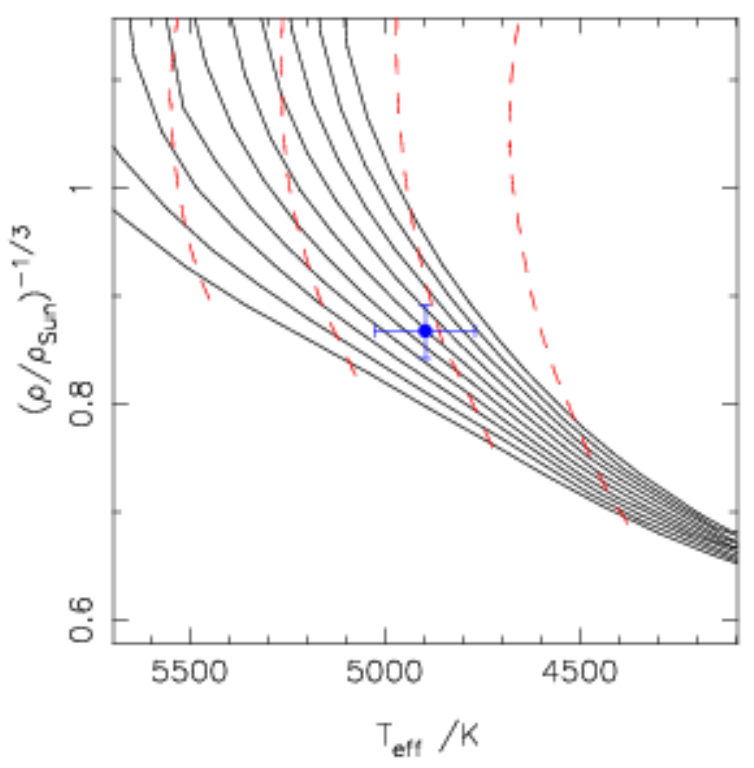}
 \caption{ The position of Qatar-1 in the $(\rho_*/\rho_\odot)^{-1/3}$ 
plane compared to theoretical evolutionary tracks and isochrones 
interpolated from \citet{2003ApJS..144..259Y} to [Fe/H]=0.2. The isochrones displayed 
represent 2.0, 4.0, 6.0, 8.0, 10.0, 12.0, 14.0, 16.0, 18.0 and 20.0~Gyr. 
The evolutionary tracks represent 0.7, 0.8, 0.9 and 1.0~$M_\odot$.
 \label{fig:modelfit}}

\end{figure}

The spectroscopic analysis of the host star Qatar-1 indicates that it is a slowly-rotating,
slightly metal-rich dwarf star of spectral type K3V. The MCMC analysis of the transit 
duration and impact parameter yields a direct estimate of the stellar density.
When compared with evolutionary tracks and isochrones for [Fe/H]=0.2 in the 
$(\rho_\star/\rho_\odot)^{1/3}$ versus $T_{\rm eff}$ plane (Fig.~\ref{fig:modelfit}), the spectroscopically-measured effective temperature indicates a mass
between 0.76 and 0.87 $M_\odot$, in good agreement with the MCMC estimate using the 
calibration of \citet{2010A&A...516A..33E}. 
The stellar density derived from the MCMC analysis
is substantially lower than would be expected for a star of this age on the zero-age main 
sequence, giving a lower limit on the stellar age of 4 Gyr. The slow 
stellar rotation rate derived from the spectra is also consistent with a spin-down age $>4$ Gyr.

The planet Qatar-1b is 10\%\ more massive than Jupiter, and has a radius 16\%\
greater than Jupiter's. It orbits its primary every 34 hours, making it one of the shortest-period
planets yet found orbiting a star less massive than the Sun. The blackbody equilibrium 
temperature given in Table~\ref{tab:mcmc} is calculated assuming a planetary albedo of 
zero and isotropic re-radiation of the power received from the host star.  A more general 
estimate of the dayside temperature is $T_{\rm eql}^4=T_\star^4(R_\star/2a)^2 ((1-A)/F)$, where $A$ 
is the planet's Bond albedo and $F$ is the fraction of the stellar surface that re-radiates 
at $T_{\rm eql}$. Measurements of starlight reflected at optical wavelengths
from the dayside hemispheres of the hot Jupiters HD 209458b \citep{2008ApJ...689.1345R}  and 
CoRoT-2b \citep{2009A&A...501L..23A} indicate that their albedos are typically of order a few 
percent. The blackbody value of 1399 K thus constitutes a reasonable lower limit on the 
planet's dayside temperature.

Qatar-1b was one of the very first batch of transit candidates from the Alsubai Project to be subjected to spectroscopic reconnaissance and radial-velocity follow-up. The rapidity of the discovery, and the fact that the planet orbits a mid-K dwarf, confirm that the instrument is well-suited to the efficient discovery of planets around lower main-sequence stars.
 
\bibliographystyle{mn2e}

\bsp

\label{lastpage}


\end{document}